\documentclass[aps,prl,twocolumn,nofootinbib]{revtex4}
\usepackage{epsfig}
\usepackage{bm}
\usepackage{amssymb}
\usepackage{amsmath}
\usepackage{color}
\usepackage{soul}
\usepackage{subfigure}
\usepackage[normalem]{ulem}

 \setstcolor{red}

\begin{document}

\title{Thermal photon radiation in high multiplicity p+Pb collisions at the Large Hadron Collider}
\author{C.\ Shen}
\affiliation{Department of Physics, McGill University, 3600 University Street, Montreal,
QC, H3A 2T8, Canada}
\author{J.-F.\ Paquet}
\affiliation{Department of Physics, McGill University, 3600 University Street, Montreal,
QC, H3A 2T8, Canada}
\author{G.\ S.\ Denicol}
\affiliation{Department of Physics, McGill University, 3600 University Street, Montreal,
QC, H3A 2T8, Canada}
\author{S.\ Jeon}
\affiliation{Department of Physics, McGill University, 3600 University Street, Montreal,
QC, H3A 2T8, Canada}
\author{C.\ Gale}
\affiliation{Department of Physics, McGill University, 3600 University Street, Montreal,
QC, H3A 2T8, Canada}

\begin{abstract}
The collective behaviour of hadronic particles has been observed in high multiplicity proton-lead collisions
at the Large Hadron Collider (LHC),  as well as in deuteron-gold collisions at
the Relativistic Heavy-Ion Collider (RHIC). In this work we present the
first calculation, in the hydrodynamic framework, of thermal photon
radiation from such small collision systems.  Owing to their compact size, these systems can reach
temperatures comparable to those in central nucleus-nucleus collisions. The thermal photons
can thus shine over the prompt background, and increase the low $p_T$ direct photon spectrum by a factor of 2-3 in 0-1\% p+Pb collisions at 5.02 TeV. This thermal
photon enhancement can therefore serve as a clean signature of the existence of
a hot quark-gluon plasma during the evolution of these small collision systems, as well as validate hydrodynamic behavior in small systems. 
\end{abstract}

\pacs{12.38.Mh, 47.75.+f, 47.10.ad, 11.25.Hf}
\maketitle
\date{\today }

\textit{1. Introduction}. The experimental heavy-ion collision program
conducted at the Relativistic Heavy-Ion Collider (RHIC) and the Large Hadron
Collider (LHC) aims to create and study the Quark-Gluon Plasma (QGP), a new
phase of nuclear matter. Relativistic hydrodynamics is the
standard theoretical framework used to describe the dynamical evolution of QGP created in
ultrarelativistic heavy ion collisions; fluid-dynamical modeling has  been very successful in describing a wide variety of measurements made at RHIC and the LHC. 
Through combined theoretical and experimental efforts, it was shown that the QGP
produced at RHIC and LHC behaves as a strongly coupled fluid, with one of
the smallest shear viscosity to entropy density ratio ever observed \cite%
{Gale:2013da,Heinz:2013th}. 

Nevertheless, the applicability of relativistic hydrodynamics has its limits; 
this theory is only valid in systems where the separation
between microscopic and macroscopic distance/time scales is sufficiently
large. This is expected to be the case in central through mid-peripheral
ultrarelativistic heavy ion collisions, where the energies reached are high
enough to produce the QGP and the volume is large enough to ensure that the
system approaches the thermodynamic limit. However, it is not clear that
this will be the case in proton-nucleus and proton-proton collisions, where
QCD matter is produced in considerably smaller volumes.

Recently, signatures usually associated with hydrodynamic behavior have also
been observed in high multiplicity p+Pb collisions at the LHC \cite%
{Abelev:2012ola, CMS:2012qk, Aad:2012gla} and d+Au collisions at the RHIC 
\cite{Adare:2013piz}. In particular, multi-particle correlations among the
produced hadrons (usually associated with collectivity) \cite%
{Khachatryan:2015waa} and mass ordering of the identified particle elliptic
flow coefficient (usually associated with radial flow) \cite%
{ABELEV:2013wsa,Khachatryan:2014jra,Adare:2014keg}, were observed 
in the p+Pb collisions
at the LHC. This came as a surprise, as such systems were believed to
be too small to produce a strongly interacting fluid. The possibility that
the QGP can also be produced in such small systems is currently a topic of
intense debate in the field.

However, even though the above signals strongly support the
fluid-dynamical nature of high multiplicity p+Pb and d+Au
collisions, they do not yet represent concrete proof. 
In order to reach more concrete conclusions, one must first
disentangle the initial-state \cite{Dumitru:2010iy,Dusling:2012cg,Dusling:2013oia,Dumitru:2014yza} and
final-state effects \cite%
{Nagle:2013lja,Bozek:2011if,Bozek:2012gr,Schenke:2014zha,Werner:2013ipa,Kozlov:2014fqa}
in the observed collective phenomena -- something that poses a great
challenge from both the theoretical and experimental points of view.

Electromagnetic (EM) radiation from the QCD matter
created in heavy-ion collisions are recognized as clean penetrating probes 
\cite{Gale:2009gc}
and can help clarify this situation. Photons suffer
negligible final state interaction once they are produced and therefore carry valuable
dynamical information from their point of emission. At low transverse momentum, a direct photon signal that is much
larger than the expected prompt photon background has been measured by
PHENIX \cite{Adare:2008ab} in 200 GeV Au+Au collisions and by the ALICE
collaboration in 2760 GeV Pb+Pb collisions \cite{Wilde:2012wc}. The direct
photons were found to have an elliptic flow ($v_{2}$) \cite%
{Adare:2011zr,Adare:2014fwh,Bannier:2014bja,Lohner:2012ct} as large as 
that of pions. These measurements have stimulated considerable theoretical
effort in photon rate calculations \cite{Dion:2011pp,Ghiglieri:2013gia,Shen:2014nfa,Heffernan:2014mla,Gale:2014dfa} as well as in setting stricter 
constraints on the dynamical description of the medium evolution in
heavy-ion collisions \cite%
{vanHees:2011vb,vanHees:2014ida,Chatterjee:2011dw,Chatterjee:2013naa,Linnyk:2013hta,Linnyk:2013wma,Shen:2014lpa,Shen:2013vja,Shen:2013cca,Linnyk:2015tha}%
.

In this paper we calculate for the first time the thermal photon radiation
of a small and rapidly expanding QGP droplet. We find a significant yield of
direct photons originating from thermal production in high multiplicity p+Pb
collisions, which can serve as a clean signal of the existence of a hot QGP
medium in these collisions. We also consider measurements in minimum bias
p+Pb and d+Au collisions at LHC and RHIC and show that even in these cases
one can see a sizeable signal due to thermal radiation. Finally, we calculate
the anisotropic flow $v_{2,3}\{\mathrm{SP}\}$ of direct photons in high
multiplicity p+A collisions and find that it is of the same magnitude as the
one calculated in central Pb+Pb collisions. The measurement of the direct
photon $v_{2,3}\{\mathrm{SP}\}$ can further constrain the dynamical
evolution of these small systems and helps us to extract the transport
properties of a QGP droplet.

\textit{2. Model and calculation}. The event-by-event simulations employ the
public code package \texttt{iEBE-VISHNU} \cite{Shen:2014vra}, with initial
conditions generated from the Monte-Carlo Glauber (MCGlb) model. In this 
implementation of MCGlb, the entropy in the transverse plane is distributed
by summing over 2D Gaussian profiles centered at the location of each
participant. The width of the Gaussian profiles is $r =\sqrt{\sigma_\mathrm{NN}/8\pi }$, 
where $\sigma_\mathrm{NN}$ is the nucleon-nucleon inelastic cross
section, and the amount of entropy deposited by each participant fluctuates
according to a Gamma distribution. The overall entropy normalization is
fixed by fitting the observed multiplicity of charged hadrons at
midrapidity. Further details of the hydrodynamic model employed as well as
of the MCGlb model can be found in Ref. \cite{Shen:2014vra}. Note that
this prescription can describe the multiplicity distribution of charged
hadrons measured in the p+Pb collisions at 5.02 TeV \cite{Chatrchyan:2013nka}%
. 
\begin{table*}[t]
\centering
\begin{tabular}{|c|c|c|c|c|c|}
\hline
p+Pb @ 5.02 TeV & $\langle N_\mathrm{coll} \rangle$ & $\frac{dN^\mathrm{ch}}{%
d\eta}\left\vert_{\vert \eta \vert < 0.5} \right.$ & $\langle p_T
\rangle(\pi^+)$ (GeV) & $v^\mathrm{ch}_2\{2\}$ & $v^\mathrm{ch}_3\{2\}$ \\ 
\hline
0-1\% & 15.4 $\pm$ 0.03 & 57.6 $\pm$ 0.3 & 0.59 $\pm$ 0.01 & 0.056 $\pm$
0.001 & 0.018 $\pm$ 0.001 \\ \hline
0-100\% & 6.6 $\pm$ 0.01 & 16.6 $\pm$ 0.3 & 0.51 $\pm$ 0.02 & 0.034 $\pm$
0.001 & 0.007 $\pm$ 0.001 \\ \hline\hline
d+Au @ 200 GeV & $\langle N_\mathrm{coll} \rangle$ & $\frac{dN^\mathrm{ch}}{%
d\eta}\left\vert_{\vert \eta \vert < 0.5} \right.$ & $\langle p_T \rangle
(\pi^+)$ (GeV) & $v^\mathrm{ch}_2\{2\}$ & $v^\mathrm{ch}_3\{2\}$ \\ \hline
0-100\% & 8.05 $\pm$ 0.01 & 8.85 $\pm$ 0.19 & 0.46 $\pm$ 0.02 & 0.025 $\pm$
0.001 & 0.003 $\pm$ 0.001 \\ \hline
\end{tabular}%
\caption{Global hadronic observables in p+Pb collisions at 5.02 TeV and d+Au
collisions at 200 GeV. The charged hadron anisotropic flow coefficients $%
v_{2,3}^{\mathrm{ch}}\{2\}$ are integrated from 0.3 to 3.0 GeV. The number
of binary collisions within the given centrality bin are estimated using
the Monte-Carlo Glauber model.}
\label{table1}
\end{table*}
Realistic distributions of nucleons are used when sampling nuclei. The
positions of the proton and neutron composing the deuteron are sampled using
the Hulthen wavefunction \cite{hulthenfun}, as outlined in Refs. \cite{Alver:2008aq,Adler:2007aa}. For
Au and Pb nuclei, the configurations calculated in Refs. \cite{Alvioli:2009ab,Alvioli:2011sk} are used as input. 

The produced initial density profiles are evolved using viscous (2+1)-d
hydrodynamic code \texttt{VISH2+1} \cite{Song:2007ux} with the lattice-based
equation of state (EoS) \texttt{s95p-v0-PCE165} \cite{Huovinen:2009yb}.
A constant shear viscosity to entropy ratio $\eta /s=0.08$ for $T>180$ MeV is assumed.
For $T<180$ MeV, we use the parameterization \cite{Niemi:2011ix}: 
\begin{equation}
\frac{\eta }{s}(T)=0.681-0.0594\frac{T}{T_{c}}-0.544\left( \frac{T}{T_{c}}%
\right) ^{2}.
\end{equation}%
Any pre-equilibrium dynamics is neglected, and it is assumed that the fluid is at rest
 in the transverse plane when the hydrodynamic evolution begins at $\tau _{0}=0.6$ fm. 
Owing to  the fireball compact size, proton-nucleus collisions have larger initial
pressure gradients than A+A collisions. Those gradients drive 
a large expansion rate, and consequently lead to a strong hydrodynamic radial flow. 
\begin{figure}[h]
\centering
\includegraphics[width=1.0\linewidth]{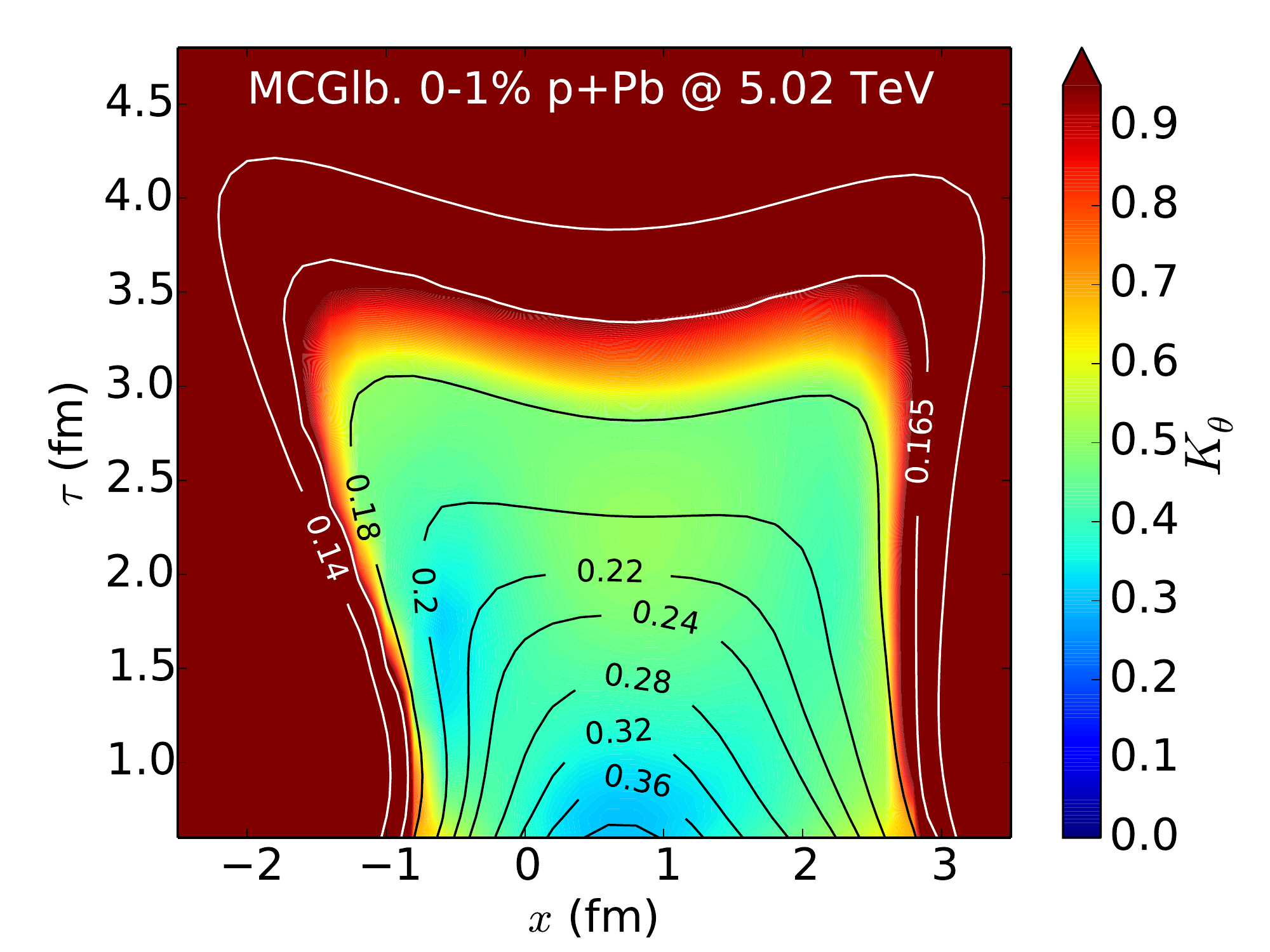}  
\caption{Color contour plot for the space-time evolution of the Knudsen
number, $K_{\protect\theta }=\protect\tau _{\protect\pi }\protect\theta $,
in 0-1\% p+Pb collisions at 5.02 TeV. Constant temperature contours are
shown. }
\label{fig1}
\end{figure}

The applicability of fluid dynamics can be characterized by the Knudsen number \cite{Niemi:2014wta},
\begin{equation}
K_{\theta }\equiv \tau _{\pi }\theta =\frac{5\eta }{e+\mathcal{P}}(\partial
\cdot u),  \label{eq.KnudsenNumber}
\end{equation}%
with small values of Knudsen number ($K_{\theta } \ll 1$) supporting the validity of a
hydrodynamic description.
Figure~\ref{fig1} shows the Knudsen number of a typical ultra-central proton-nucleus collision.
Its hydrodynamic description gradually breaks down as the temperatures
decreases and, one can see that, for temperatures $T \lesssim 0.165$ MeV, the Knudsen number is already
above one in almost all space-time points. For this reason, a kinetic freeze-out temperature $T_{\mathrm{dec}}=165$ MeV is used for the calculations that yield the results on which we report here. 
The thermal photon radiation is computed from the medium only above $T_{%
\mathrm{dec}}$. 

In the QGP phase, we use the full leading order $O(\alpha
_{s}\alpha _{\mathrm{EM}})$ photon production rate \cite{Arnold:2001ms}. In
the hadron gas phase, photon production from mesonic channels \cite{Turbide:2003si}, $\rho $
spectral function, and $\pi +\pi $ bremsstrahlung \cite{Heffernan:2014mla}
are taken into account. Shear viscous corrections are included in the 2 to 2
scattering processes in the QGP phase \cite{Shen:2014nfa} and in all the
mesonic reactions in hadron gas phase \cite{Dion:2011pp}. We switch rates from QGP to
hadron gas at $T=180$ MeV, where the equilibrium rates from both
phases are approximately the same.

The emitted thermal photon momentum distribution is computed by folding the
thermal photon production rates, $q\frac{dR}{d^{3}q}(q,T)$, with the
dynamically evolving medium, event-by-event \cite{Shen:2013cca}: 
\begin{equation}
E\frac{dN^{\mathrm{th,}\gamma }}{d^{3}p}=\int d^{4}x\bigg(q\frac{dR}{d^{3}q}%
(q,T(x))\bigg)\bigg\vert_{q=p\cdot u(x)}.  \label{eq1}
\end{equation}%
The anisotropic flow coefficients of the direct photons are calculated by
correlating them with all charged hadrons \cite{Shen:2014lpa}, 
\begin{eqnarray}
v_{n}\{\mathrm{SP}\}(p_{T}) = \frac{\left\langle \frac{dN^{\gamma }}{dp_{T}}v_{n}^{\gamma
}(p_{T})v_{n}^{\mathrm{ch}}\cos [n(\Psi _{n}^{\gamma }(p_{T})-\Psi _{n}^{%
\mathrm{ch}})]\right\rangle }{\left\langle \frac{dN^{\gamma }}{dp_{T}}%
\right\rangle v_{n}^{\mathrm{ch}}\{2\}}.  \label{eq2}
\end{eqnarray}

%
\begin{figure*}[ht!]
\centering
\begin{tabular}{cc}
\includegraphics[width=0.45%
\linewidth]{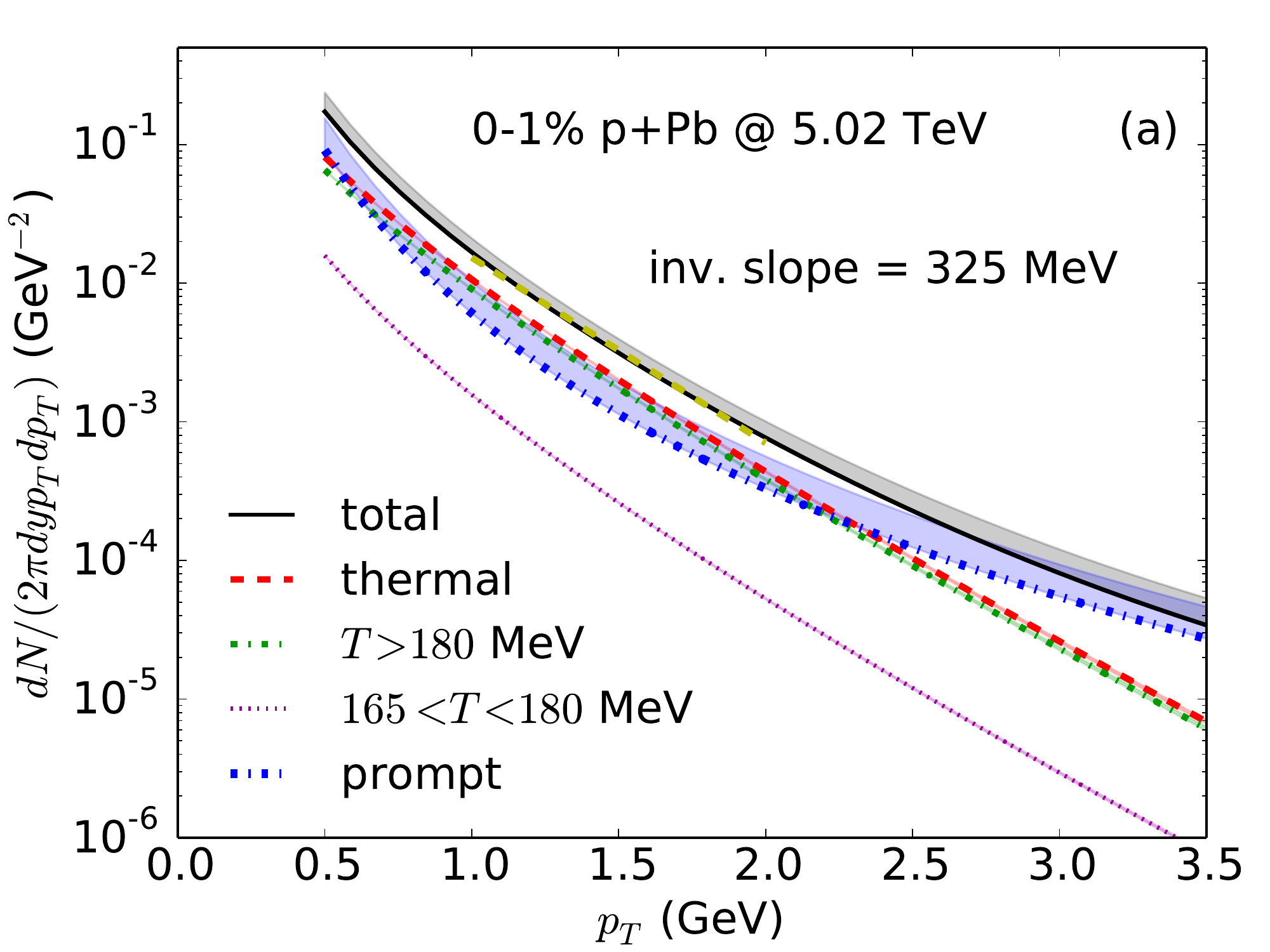} & %
\includegraphics[width=0.45%
\linewidth]{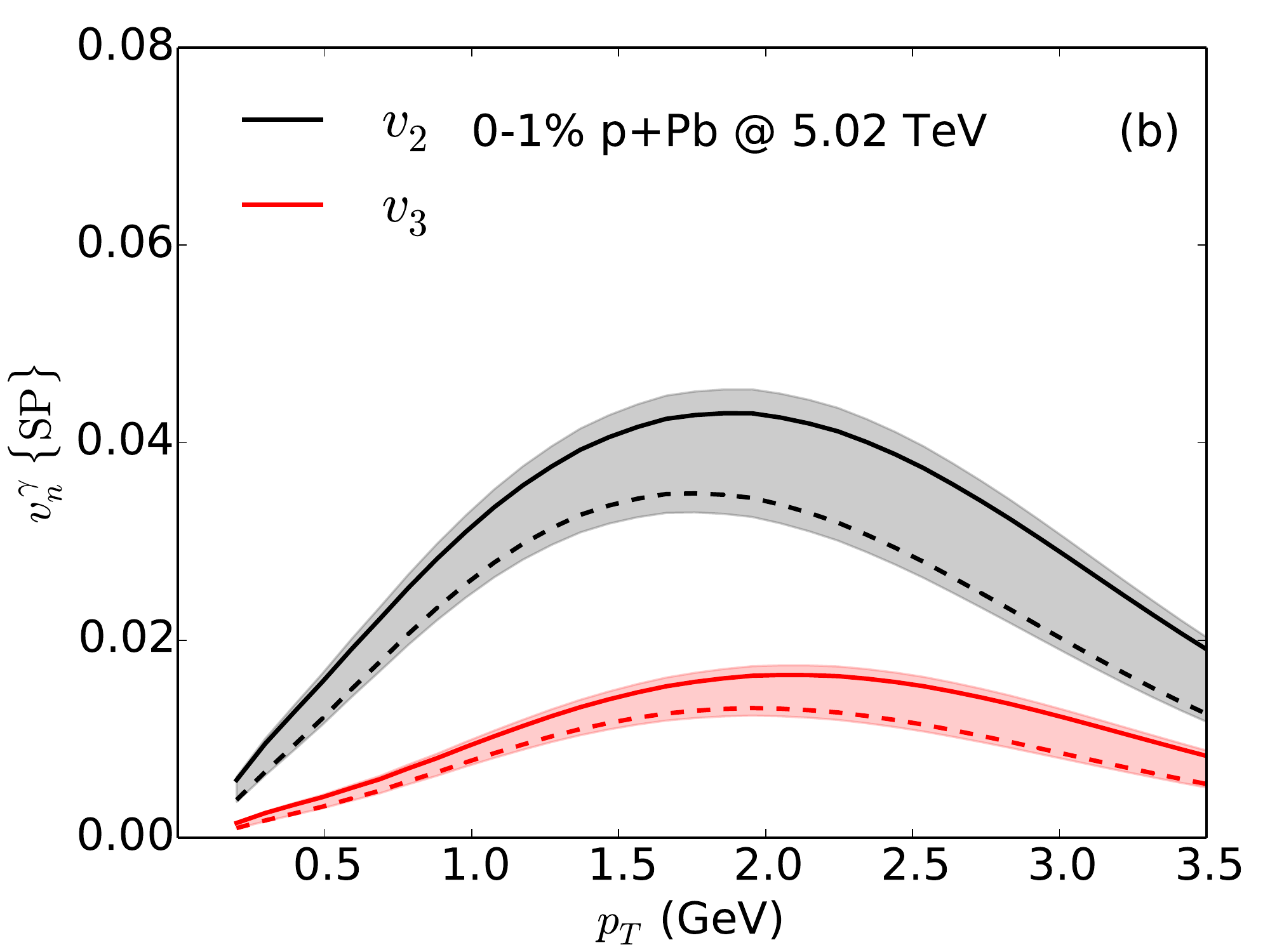}%
\end{tabular}
\caption{(a) Direct photon spectra and (b) $v_n^\gamma\{\mathrm{SP}\}$ from 0-1\% p+Pb
collisions at $\protect\sqrt{s} = 5.02$ TeV. The shaded bands indicate the theoretical uncertainty in determining $N_\mathrm{coll}$ from the MCGlb model as explained in the main text. In panel (a), the inverse slope of the direct photon spectra is obtained from fitting the inverse slope in the range $p_T \in [1.0, 2.0]$ GeV. }
\label{fig2}
\end{figure*}
%

Table \ref{table1} summarizes the global hadronic observables from our
simulations for p+Pb collisions at 5.02 TeV and d+Au collisions at 200 GeV. 
The MCGlb model with multiplicity fluctuations can correctly reproduce the
centrality dependence of charged hadron multiplicity. For p+Pb collisions,
the mean-$p_{T}$ of pions calculated is around 10\% higher than the values
experimentally observed \cite{Chatrchyan:2013eya}. The inclusion of bulk
viscosity may reduce this tension with the data \cite{Ryu:2015vwa}. The charged hadron
anisotropic flow coefficients, $v_{2,3}\{2\}$ for p+Pb collisions is in
reasonable agreement with the experimental measurements from the CMS
collaboration \cite{Chatrchyan:2013nka}, with $v_{2}\{2\}$ being slightly
underestimated by the calculation.

\textit{3. Results and Discussion}. Whether thermal photons can be observed
or not depends on their contribution relative to prompt photons. In this
letter the prompt photon background is evaluated with perturbative QCD
(pQCD) at next-to-leading order (NLO)~\cite{Aversa:1988vb,Aurenche:1987fs}, scaled with the number of binary collisions as computed in the Glauber model.
For nucleus collisions, cold nuclear effects are included by using EPS09
nuclear parton distribution functions~\cite{Eskola:2009uj}. The isospin effect
is included as well. The proton parton distribution functions used is
CTEQ61m~\cite{Stump:2003yu} and the photon fragmentation function is BFG-2~
\cite{Bourhis:1997yu}. The factorization, renormalization and fragmentation
scales are set equal to $\alpha p_{T}$, where $p_{T}$ is the transverse
momentum of the produced photon. The constant $\alpha $ is set to $1/2$ so
as to provide a good description of the available direct photon measurements
at RHIC~\cite{MScThesisJF}. The minimum scale $Q$ parametrized in parton
distribution function and fragmentation function is around $1.5$~GeV, which
for the present choice of $\alpha $ limits the pQCD calculation to $p_{T}>3$%
~GeV. Nevertheless the effect of $Q$ is predominantly a change in
normalization of the pQCD prediction, and a larger value of $\alpha $ can be
used to extrapolate the pQCD calculation to low $p_{T}$. We have verified  that
this approaches provides a reasonable description of the available low $p_{T}
$ photon data from RHIC~\cite{Adare:2012vn}, which are available down to $%
p_{T}\approx 1$ GeV. This extrapolation scheme is thus used to estimate the
prompt photon signal at low $p_{T}$ in p+p and p+A collisions.

Figure  \ref{fig2} shows the direct photon spectra and $p_{T}$%
-differential elliptic and triangular flow coefficients in 0-1\% p+Pb
collisions at 5.02 TeV. In order to calculate the prompt photon spectrum, the number of binary collisions $N_{\mathrm{coll}}$ is required. Since binary scaling is known to be difficult to estimate for ultracentral p+A events, we consider two values of $N_{\mathrm{coll}}$ and define their difference as the uncertainty inherent in the prompt photon calculation. One value of $N_{\mathrm{coll}}$ is calculated
with our MCGlb model (listed in Table \ref{table1}) and a larger one, $N_{%
\mathrm{coll}} = 26.1$, is estimated using the Glauber-Gribov model (with $\Omega = 1.01$) from the ATLAS
collaboration \cite{TheATLAScollaboration:2013cja}. Note that high-$p_T$ photon measurements for ultracentral events could help to reduce the uncertainty in $N_{\mathrm{coll}}$. Using this prescription for prompt photons, one finds that thermal
photons emitted from the hot and dense medium outshine the prompt photon
background by a factor of 2-3, for $p_T^\gamma \leq 2.5$ GeV.  We reiterate that this result is obtained by only considering
thermal photons emitted above $T=165$ MeV: If radiation emission below this
temperature were also included, the thermal enhancement would be even
larger.

Two factors explain the large thermal photon signal in these
small systems. First, the temperatures reached in the most central
p+Pb collisions is
considerably higher than the one reached in peripheral Pb+Pb collisions, and
is comparable to, or even larger than, the temperatures reached in central Pb+Pb
collisions. Second, the number of binary collisions 
in  p+Pb collisions is
much smaller than the one in Pb+Pb collisions. This last factor 
reduces the background signal from the prompt photon component, while the first 
 increases the thermal photon production.

Importantly, the significant blue shift from hydrodynamic radial flow and the high
temperatures reached at the early stages of the collision result in an inverse slope of 325 MeV for the direct photon spectrum, which is harder than the one measured in 0-40\%  Pb+Pb collisions \cite{Wilde:2012wc}.

Fig. \ref{fig2}b shows the scalar-product anisotropic flow coefficients of
direct photons in p+Pb collisions. The dashed lines represent results obtained 
with the prompt photons estimated with the $N_{\mathrm{coll}}$ from the
ATLAS Glauber-Gribov model \cite{TheATLAScollaboration:2013cja}. The bands represent the uncertainty in prompt photon production discussed previously, and also include the statistical error in thermal photon production from a finite number of hydrodynamical calculations. 
Unlike the situation in nucleus-nucleus collisions, in p+Pb events the direct photon
anisotropic flow is seeded by the density fluctuations of the
initial state. We find that the direct photon $v_{2,3}\{\mathrm{SP}\}(p_{T})$
have roughly the same sizes compared to 0-40\% centrality in Pb+Pb collisions 
\cite{Shen:2013cca}. Because this large anisotropy of direct photons is 
generated during the collective expansion of the fireball, the measurement
of photon flow observables can provide an independent validation of hydrodynamics in environments with small volumes and large pressure gradients.  A
global  analysis with hadronic observables can therefore lead to tight 
constraints on the transport properties of the QGP.

\begin{figure}[h!]
\centering
\includegraphics[width=1.0\linewidth]{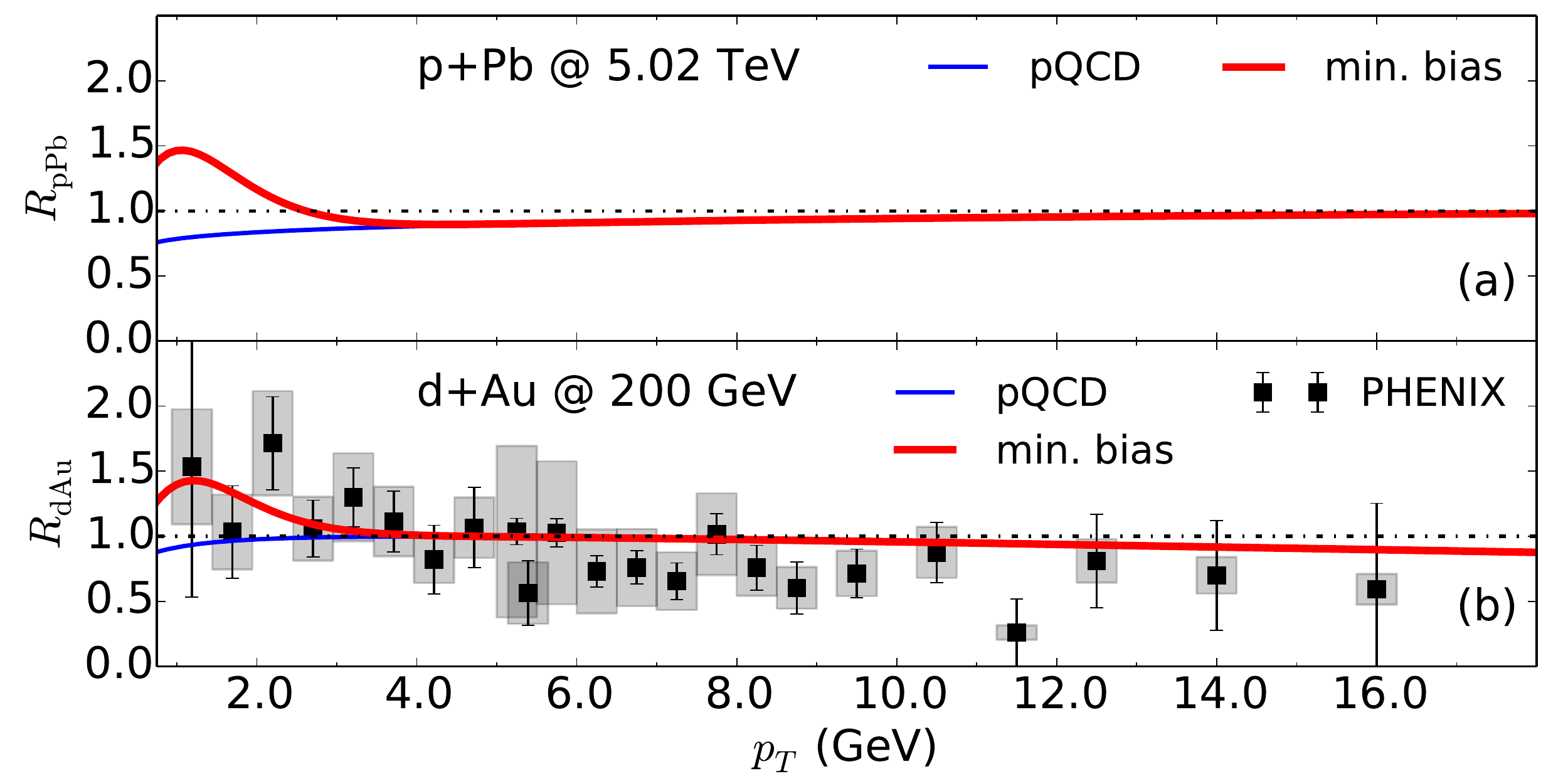}  
\caption{The nuclear modification factor for direct photons in minimum bias
p+Pb collisions at 5.02 TeV and d+Au collisions at 200 GeV \protect\cite%
{Adare:2012vn}. The calculations of prompt photon emission include isospin and shadowing corrections.}
\label{fig3}
\end{figure}
%
As photon production in  highly central collisions of light-heavy ions has yet to be measured, we estimate the thermal component in the nuclear
modification factor $R_\mathrm{pPb}$ and $R_\mathrm{dAu}$ for direct photons
in minimum bias collisions in Fig. \ref{fig3}, and compare with existing data at RHIC.
By including the thermal radiation component, we find a sizeable enhancement
of $R_{\mathrm{pPb}}$ and $R_{\mathrm{dAu}}$ over the prompt baseline for $%
p_{T}<3$ GeV. The thermal photon radiation leaves a clear and robust
measurable signal in the minimum bias measurement. Our results for d+Au
collisions at 200 GeV are consistent with the current PHENIX measurements 
\cite{Adare:2012vn}. A reduction of the uncertainties in the experimental
data at the low $p_{T}$ has the potential to distinguish the pure prompt
production scenario from one with additional thermal radiation. For p+Pb
collisions at 5.02 TeV, the signal of the thermal enhancement is more pronounced
than the one at RHIC energy. The nuclear modification factor is however roughly the same, owing to a smaller  prompt photon $R_\mathrm{pA}$ at the LHC. The observation of such an enhancement in
the direct photon nuclear modification factor can serve a clean signature of
the existence of quark-gluon plasma in small collision systems. 

\textit{4. Conclusions}. In this paper, we propose to use direct photons as
a signature of the existence of the hot quark-gluon plasma in the d+Au and
p+Pb collisions at the RHIC and the LHC. Owing to compact  fireball sizes, 
these systems can achieve high temperatures, comparable with those in 
central Pb+Pb collisions. Compared to A+A collisions, a smaller number of
binary collisions reduces the background of prompt photons. These two
factors cause the thermal photon signal to shine over its prompt counterpart in
high multiplicity events. 
Future work will consider three spatial dimensions \cite{Schenke:2010nt}, the presence of a semi-QGP \cite{Gale:2014dfa}, the introduction of an IP-Glasma initial state \cite{Schenke:2012wb}, a possible hard photon component \cite{McLerran:2015mda}, and the inclusion of a coefficient of bulk viscosity \cite{Ryu:2015vwa}.
We predict that the anisotropic flow
of direct photons in p+Pb collisions will be comparable to those measured in Pb+Pb
collisions. It is found that the thermal photon radiation can also  leave a clear,
measurable trace in  minimum bias d+Au and p+Pb collisions at  RHIC
and at the LHC. Precise measurements of direct photon spectra at
low $p_T$ have the potential to reveal the quark-gluon plasma formed in these
light-heavy ion collisions.

\textit{Acknowledgments}. The authors thank Friederike Bock for useful discussions.
This work was supported by the Natural Sciences and Engineering Research
Council of Canada. G.~S.~Denicol acknowledges support through a Banting
Fellowship of the Natural Sciences and Engineering Research Council of
Canada. Computations were made on the Guillimin supercomputer at McGill
University, managed by Calcul Qu\'{e}bec and Compute Canada. The operation
of this supercomputer is funded by the Canada Foundation for Innovation
(CFI), Minist\`{e}re de l'\'{E}conomie, de l'Innovation et des Exportations
du Qu\'{e}bec (MEIE), RMGA and the Fonds de recherche du Qu\'{e}bec - Nature
et technologies (FRQ-NT).

\end{document}